\documentclass[a4paper,11pt]{article}

\usepackage{pos}
\usepackage{braket}
\usepackage{hyperref}
\usepackage{graphicx}
\usepackage[utf8]{inputenc}
\usepackage{physics}
\usepackage{subcaption}
\usepackage{tikz}
\usepackage[compat=1.1.0, /tikzfeynman/warnluatex=false]{tikz-feynman}

\newcommand*{\lr}[1]{\left( {#1} \right)}
\newcommand*{\lrb}[1]{\left[ {#1} \right]}

\newcommand*{\e}{\operatorname{e}}
\newcommand*{\im}{\operatorname{i}}

\title{Semileptonic $D \rightarrow \pi \ell \nu$, $D \rightarrow K \ell \nu$ and $D_s \rightarrow K \ell \nu$ decays with 2+1f domain wall fermions}
\ShortTitle{Semileptonic $D_{\left(s\right)} \rightarrow K / \pi \ell \nu$ decays with 2+1f domain wall fermions}

\author[a, b]{Peter Boyle}
\author[b]{Luigi Del Debbio}
\author[b]{Felix Erben}
\author[c]{Jonathan Flynn}
\author[c, d]{Andreas Jüttner}
\author*[b]{Michael Marshall}
\author[b]{Antonin~Portelli}
\author[e]{J~Tobias~Tsang}
\author[f]{Oliver Witzel}

\affiliation[a]{Brookhaven National Laboratory, Upton, NY 11973, USA}

\affiliation[b]{Higgs Centre for Theoretical Physics, School of Physics and Astronomy, The University of Edinburgh,\\
James Clerk Maxwell Building, Peter Guthrie Tait Road, Edinburgh, EH9 3FD, UK}

\affiliation[c]{Physics and Astronomy, University of Southampton, Southampton, SO17 1BJ, UK}

\affiliation[d]{ Theoretical Physics Department, CERN, 1211 Geneva 23, Switzerland}

\affiliation[e]{CP3-Origins and IMADA, University of Southern Denmark, Campusvej 55, DK-5230 Odense M, Denmark}

\affiliation[f]{Center for Particle Physics Siegen, Theoretische Physik 1, Naturwissenschaftlich-Technische Fakult{\"a}t, Universit{\"a}t Siegen, 57068 Siegen, Germany}

\emailAdd{Michael.Marshall@ed.ac.uk}

\abstract{We present the status of our project to calculate $D \to \pi \ell \nu$, $D \to K \ell \nu$ and $D_s \to K \ell \nu$ semileptonic form factors using domain wall fermions for both heavy and light quarks. Our computations are performed using RBC/UKQCD's set of 2+1 flavour domain wall fermion and Iwasaki gauge field ensembles. We plan to calculate three-point functions covering the full, physically allowed kinematic range. Given that the signal decays faster than the noise, unambiguously and reliably extracting the ground state is critical for success. We include an analysis of operator diagonalisation within several possible $2 \times 2$ operator bases and find an admixture of gauged fixed wall and $\mathbb{Z} \left( 2 \right)$ wall sources to be acceptable at both zero and non-zero momentum.  Initial results for semileptonic form factors are presented for first ensembles.}

\FullConference{%
 The 38th International Symposium on Lattice Field Theory, LATTICE2021
  26th-30th July, 2021
  Zoom/Gather@Massachusetts Institute of Technology
}

\begin{document}
\maketitle


\section{Introduction}

The six-quark model and the Cabibbo-Kobayashi-Maskawa (CKM) matrix was first proposed as a ``very interesting and elegant'' \cite{Kobayashi:2009, cabibbo:1963, Kobayashi:1973} mechanism to explain CP-violation. Half a century later, the model still stands and the study of flavour-changing processes has become the field of flavour physics.

Precise experimental measurements from charm factories such as CLEO-c and BESIII and bottom-factories such as Belle, Belle II, BaBar and LHCb continue to test the Standard Model and the unitarity of the CKM matrix to ever greater precision. Hints of potential new physics exist but to resolve them requires increased precision in theoretical prediction and experimental results.

\begin{figure}
\begin{center}
\begin{tikzpicture}[scale=01, every node/.style={transform shape}]
\begin{feynman}[large]
\vertex [blob, fill=red!20!white, style={/tikz/minimum size=1.25cm}] (lb) {$D_{\lr{s}}$};
\vertex [blob, fill=blue!20!white, style={/tikz/minimum size=1.25cm}, right=50mm of lb] (rb) {$K / \pi$};
\vertex (l1) {};
\vertex [right=4.75mm of lb, yshift=2.5mm] (l1) {};
\vertex [below=5mm of l1] (l2) {};
\vertex [left=4.75mm of rb, yshift=2.5mm] (r1) {};
\vertex [below=5mm of r1] (r2) {};
\vertex [right=20mm of l1] (w1);
\vertex [right=5mm of w1, yshift=10mm] (w2);
\vertex [right=12.5mm of w2, yshift=-2mm] (nu) {$\nu_\ell$};
\vertex [above=6mm of nu] (lep) {$\ell$\,\,};
\diagram* {
(l1) -- [fermion, edge label=$c$] (w1) [square dot] -- [fermion, edge label=$d/s$] (r1),
(r2) -- [fermion, edge label=$s/d$] (l2),
(w1) -- [photon, edge label=$W$] (w2) -- [fermion] (nu),
(lep) -- [fermion] (w2),
};
\draw [decoration={brace}, decorate] (lep.north east) -- (nu.south east) node [pos=0.25, right] {$q = p_i - p_f$};
\draw node [left=9.5mm of lb] {$p_i$};
\draw node [right=10mm of rb] {$p_f$};
\end{feynman}
\end{tikzpicture}
\end{center}
\caption{Tree-level semileptonic decay of an initial pseudoscalar $D_{\lr{s}}$ meson (with momentum $p_i$) to a final pseudoscalar $K$ or $\pi$ meson (momentum $p_f$). Momentum $q = p_i - p_f$ is transferred to the final-state $\ell \nu$ pair.} \label{fig:continuum_decay}
\end{figure}
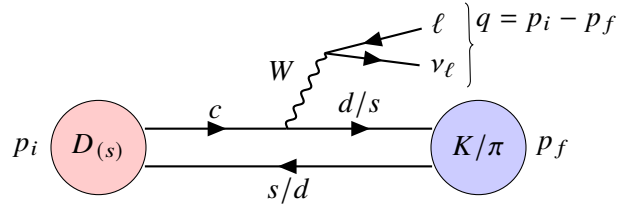

The aim of this work is to perform lattice computations of the matrix elements of exclusive semileptonic meson decays involving $c \rightarrow d$ and $c \rightarrow s$ flavour transitions (fig~\ref{fig:continuum_decay}). We use this to extract the $q^2$-dependence of the relevant form factors over the entire physically allowed kinematic range.

Experimental results for semileptonic meson decays quote products of form factors and CKM matrix elements. Recent HFLAV values for $D$-meson decays \cite{HFLAV:2019otj, Ablikim:2017, Ablikim:2019} give $\abs{V_{cs} f_+^{D \to K} \lr{0}} = 0.7133(68)$ and $\abs{V_{cd} f_+^{D \to \pi} \lr{0}} = 0.1400(33)$, i.e. errors of 1\% and 2.4\% respectively. By combining the experimental results with our lattice form factors, we aim to extract $\abs{V_{cd}}$ and $\abs{V_{cs}}$ which will allow us to test the unitarity of the CKM matrix in the Standard Model.
 
Our immediate goal is to perform the form factor determination to percent-level accuracy in order to be commensurate with the experimental results. Our approach complements the existing literature by performing the computation entirely with domain wall fermions. For the charm quark we utilise the discretisation employed in~\cite{Boyle:2018knm}, using a stout-smeared \cite{Morningstar:2003gk} Möbius \cite{Brower:2012vk} action. The light and strange quarks are simulated with the Shamir \cite{Kaplan:1992bt, Shamir:1993zy, Furman:1995ky, Blum:1996jf, Blum:1997mz} kernel.


\section{Point-wall diagonalisation study}

In order to extract form factors with the smallest possible variance, we seek methods to reliably eliminate excited state contamination from the two- and three-point functions we use on the earliest possible timeslice. Our earlier study of pseudoscalar-axial diagonalisation proved equivocal \cite{Boyle:2019cdl}. In this study we examined whether linear combinations of correlators constructed from point and wall operators could be used.

We label pseudoscalar mesons $\ket{P_i, n}$ where $P_i$ labels the initial or final meson ($D$, $D_{s}$, $K$ or $\pi$), $n = 0$ labels ground and $n > 0$ label excited states. We use $\ket{\Omega}$ to denote the vacuum. When labelling an operator (and anything derived from an operator), the subscript $i \equiv$ initial and $f \equiv$ final also includes $P \equiv \textrm{point}$ and $W \equiv \textrm{wall}$ variants.

This study was performed on the RBC/UKQCD C1 ensemble \cite{Aoki:2010pe}: $a^{-1} = 1.78$ GeV; $\flatfrac{L}{a} = 24$; $\flatfrac{T}{a} = 64$; $m_\pi = 340$ MeV; $m_\pi L = 4.57$. This preliminary study uses a stout-smeared \cite{Morningstar:2003gk} Shamir \cite{Kaplan:1992bt, Shamir:1993zy, Furman:1995ky, Blum:1996jf, Blum:1997mz} action for the charm, rather than stout-smeared Möbius \cite{Brower:2012vk} used in our production setup. This preliminary data set has 35 configurations, binning measurements from 16 timeslices on each.

\subsection{Two-point diagonalisation study}

We construct two-point correlation functions $C^{\lr{2}}_{if}$ (with $i, f \in \set{P,W}$, i.e. point or wall) using interpolating operators $O = \bar{\psi}_2 \Gamma \psi_1$ with $\Gamma \in \set{\gamma_5, \gamma_T \gamma_5}$ appropriate for pseudoscalar mesons. Defining $A_{f,n} \equiv \mel{\Omega}{O_f}{n}$ and using the labels $i \equiv$ initial and $f \equiv$ final, the correlation function can be parameterised
as
\begin{align}
\label{eq:corr_2pt} C^{\lr{2}}_{if} \lr{t, \vb{p}} \equiv \sum_{\vb{x}} \e^{\im \vb{p} \cdot \vb{x}} \ev{O_f \lr{t, \vb{x}} O_i^\dag \lr{0, \vb{0}}} &= \sum_n \frac{A_{f,n} A_{i,n}^*}{2 E_n} \lr{\e^{- E_n t} \pm \e^{- E_n \lr{T - t}}} . \qquad \qquad
\intertext{Consideration of the ground and first excited state in \eqref{eq:corr_2pt} leads us to define a linear combination for a source mixing angle $\theta$ (where the first excited states are expected to cancel for $\theta = - \flatfrac{\pi}{4}$)}
\label{eq:mixed_2pt} C^{\lr{2}}_{\textrm{mixed}} \lr{\theta, t, \vb{p}} &= \frac{\sin \theta}{A_{W, 1}}C^{\lr{2}}_{WP} \lr{t, \vb{p}} + \frac{\cos \theta}{A_{P, 1}}C^{\lr{2}}_{PP} \lr{t, \vb{p}},
\end{align}
where we extract the $A_{f,n}$ from simultaneous fits to the point and wall correlation functions.
\begin{figure}
\begin{center}
\includegraphics{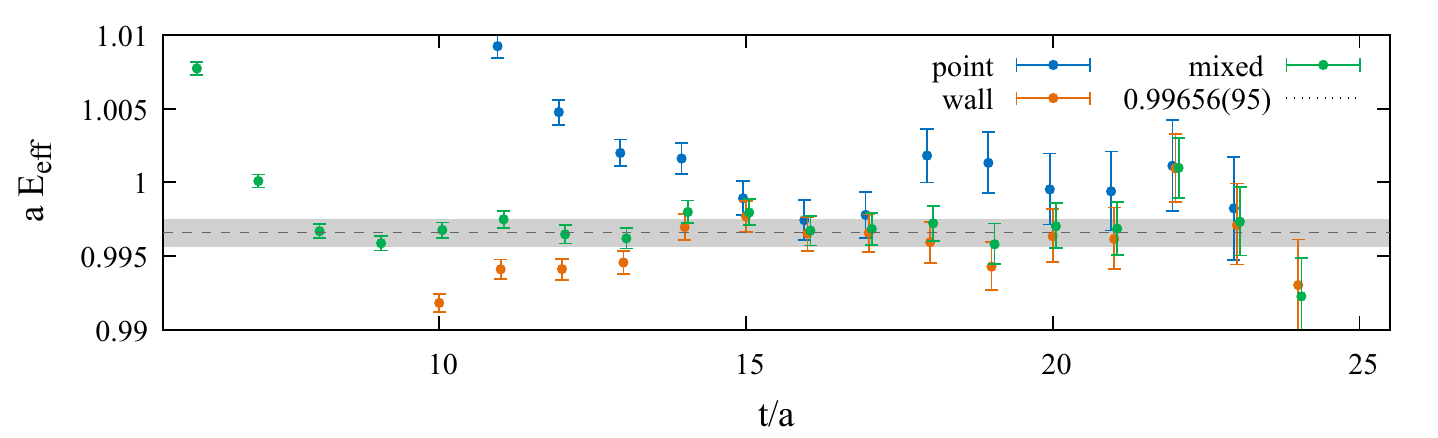}
\end{center}
\caption{Effective mass plot $a E_{\textrm{eff}} \lr{t} = \cosh^{-1} \lr{\flatfrac{C \lr{t - 1} + C \lr{t + 1}}{2 C \lr{t}}}$ for a $D$-meson $C^{\lr{2}}_{\textrm{mixed}} \lr{t, \vb{0}}$ constructed from $C_{PP}$ and $C_{WP}$ using overlap coefficients extracted from combined fit on timeslices $\lrb{7,17}$. The green data points are obtained for an angle of $\theta = - \flatfrac{\pi}{4}$. As expected, this choice leads to a significant reduction in excited state contamination. The grey dashed line (grey band) is the prior published result (variance) \protect\cite{Boyle:2018knm} (Table VI, $a m_h=0.58$)} \label{fig:cancel_2pt}
\end{figure}

We observe that the mixed correlator reaches a plateau around timeslice 8, much earlier than the underlying point and wall correlators (which plateau around timeslice 15), and the mass is compatible with prior published results \cite{Boyle:2018knm} (Table VI, $a m_h=0.58$). Plateauing much earlier, the relative error of the mixed correlator is much smaller than that of the underlying correlators when they reach plateau. We conclude that excited state cancellations occur as expected.


\subsection{Three-point diagonalisation study}

Heavy-light/strange three-point functions with local current $V_\mu$ have the form (ignoring around-the-world effects and assuming $P_i$ is located on timeslice 0) 
\begin{align}
\label{eq:three_pt} C_{if}^\mu \lr{\Delta T, t, \vb{p}_i \vb{p}_f} &= \sum_{m, n=0}^\infty \frac{A_{f, m} A_{i, n}}{4 E_{f, m} E_{i, n}} \mel{P_f \lr{\vb{p}_f}, m}{V^\mu (q^2) }{P_i \lr{\vb{p}_i}, n} \e^{- \lr{E_{i,n} - E_{f, m}} t} \e^{- E_{f,m} \Delta T}.
\intertext{We construct two symmetric (with respect to the sink and source meson) double ratios, $R^\mu_1$ and $R^\mu_2$ \cite{Flynn:2007, Boyle:2013}, which are designed to approach the renormalised ground state matrix element for $\Delta T \gg t \gg 0$, leaving the matrix element of the renormalised current}
\label{eq:R_alpha} R_\alpha^\mu \lr{\vb{p}_i, \vb{p}_f} &= 2 \sqrt{\frac{E_i E_f}{D_\alpha}} \sqrt{C^{\mu}_{if} \lr{\vb{p}_i, \vb{p}_f} \, C^{\mu}_{fi} \lr{\vb{p}_f, \vb{p}_i}}
\approx Z_V \mel{P_f \lr{\vb{p}_f}}{V^\mu (q^2)}{P_i \lr{\vb{p}_i}},
\end{align}
where $a \in \set{1, 2}$ and the denominator is: $D_1 = \flatfrac{C^{\lr{2}}_{ii} \lr{\vb{p}_i} \, C^{\lr{2}}_{ff} \lr{\vb{p}_f}}{\lr{Z_{V\textrm{h}} Z_{V\textrm{l}}}}$ (see section~\ref{sec:Z_V} for $Z_{V\textrm{h}} \equiv Z_{V \, \textrm{heavy}}$ and $Z_{V\textrm{l}} \equiv Z_{V \, \textrm{light/strange}}$); or $D_2 = C^{0}_{ii} \lr{\vb{p}_i, \vb{p}_i} \, C^{0}_{ff} \lr{\vb{p}_f, \vb{p}_f}$. We define mixed three-point functions (simplifying the notation and introducing arbitrary constants $\alpha$, $\beta$, $\gamma$ and $\delta$)
\begin{align}
C^\mu_{\substack{\textrm{source}\\\textrm{mixed}}, f} &= \alpha \, C^\mu_{\textrm{point}, f} \quad + \beta \, C^\mu_{\textrm{wall}, f} \qquad,\\
C^\mu_{\textrm{mixed}} \quad &= \gamma \, C^\mu_{\substack{\textrm{source}\\\textrm{mixed}}, \textrm{point}} + \delta \, C^\mu_{\substack{\textrm{source}\\\textrm{mixed}}, \textrm{wall}}.
\intertext{We can show that if we introduce tunable mixing angles $\phi$ at sink and $\theta$ at source}
\gamma = \frac{\cos \phi}{A_{f,1}^{\lr{\textrm{point}}}} , \qquad \delta &= \frac{\sin \phi}{A_{f,1}^{\lr{\textrm{wall}}}} , \qquad \alpha = \frac{\cos \theta}{A_{i,1}^{\lr{\textrm{point}}}} \quad \textrm{and} \quad \beta = \frac{\sin \theta}{A_{i,1}^{\lr{\textrm{wall}}}},
\end{align}
then we expect excited state cancellations. That is, near $\phi = \theta = - \flatfrac{\pi}{4}$ the mixed three-point function approaches this form
\begin{align}
\label{eq:mixed_3pt} C^\mu_{\textrm{mixed}} &\approx \frac{\lr{\gamma A_{f,0}^{\lr{\textrm{point}}} + \delta A_{f,0}^{\lr{\textrm{wall}}}} \lr{\alpha A_{i,0}^{\lr{\textrm{point}}} + \beta A_{i,0}^{\lr{\textrm{wall}}}}}{4 E_f E_i} \mel{P_f}{V^\mu}{P_i} \e^{- \lr{E_i - E_f} t} \e^{- E_f \Delta T},
\end{align}
which is the exponential behaviour of the ground-state. We extract the $A_{f/i,1}^{\lr{\textrm{point/wall}}}$ from simultaneous fits to the point and wall correlation functions, $C^{\lr{2}}_{PP} \lr{t, \vb{p}}$ and $C^{\lr{2}}_{WP} \lr{t, \vb{p}}$.
In the absence of exact knowledge of the numerical values of $A_{f/i,1}^{\lr{\textrm{point/wall}}}$, the parameters $\alpha$, $\beta$, $\gamma$ and $\delta$ can be tuned to optimise the excited state cancellation.
\begin{figure}
	\centering
	\begin{subfigure}[t]{0.47\textwidth}
		\centering
		\includegraphics{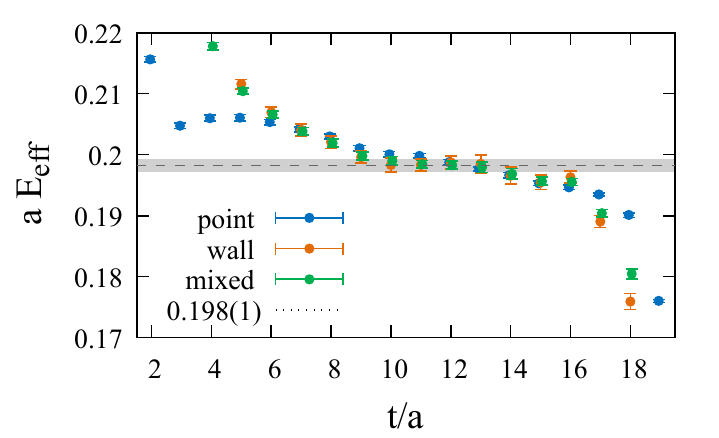}
	\end{subfigure}
	\begin{subfigure}[t]{0.47\textwidth}
		\centering
		\includegraphics{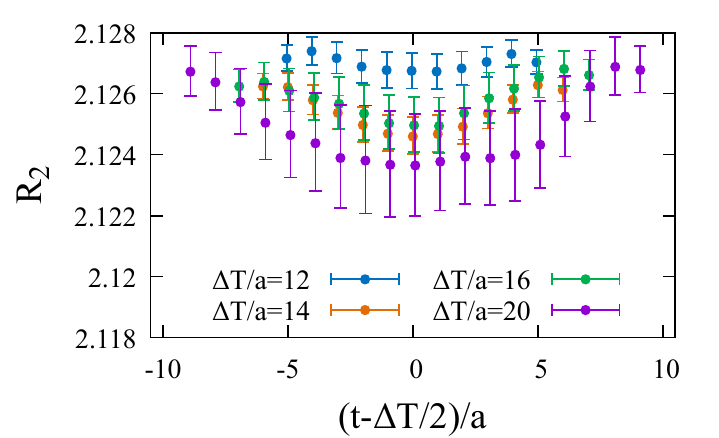}
	\end{subfigure}
	\caption{The (unphysical) transition $D_s \eval_{a m_c = 0.69} \rightarrow D_s \eval_{a m_c = 0.50}$. \emph{Left:} $C^\mu_{\textrm{source mixed}, \textrm{point}} \lr{\theta = - \frac{\pi}{4}}$, $\flatfrac{\Delta T}{a}=20$. Near the mid-point, the ground-state in \protect\eqref{eq:three_pt} dominates, and the effective mass of the three-point function approximates the meson mass difference, 0.198(1) \protect\cite{Boyle:2018knm}. \emph{Right:} Mixed $R_2$ ratio at $\theta = \phi \sim-80^\circ$ (determined from a scan over mixing angles). We see agreement for multiple wall separations $\flatfrac{\Delta T}{a}$ (except for $\flatfrac{\Delta T}{a}=12$). The statistical variance increases as the wall separation is increased.} \label{fig:3pt_study}
\end{figure}

When we plot the effective mass of the source mixed correlation function (left panel, fig~\ref{fig:3pt_study}), the correlator is an interpolation between the point and wall sources. There is no obvious improvement as there was with the two-point source mixed correlation function. Tuning the mixing angle(s) makes little difference.

Similarly, when we plot the $R_\alpha$ ratios, we see no obvious improvement near $\phi = \theta \sim -\frac{\pi}{4}$. If a scan over all mixing angles is performed, we see that the optimal mixing angle (right panel fig~\ref{fig:3pt_study}) involves mostly the wall-source.


\section{Preliminary $D_{\lr{s}} \rightarrow \flatfrac{K}{\pi} \ell \nu$ data production}

Data presented in this section were created on the RBC/UKQCD M1 ensemble \cite{RBC:2010qam, RBC-UKQCD:2008mhs}: $a^{-1} = 2.38(1)$ GeV; $\flatfrac{L}{a} = 32$; $\flatfrac{T}{a} = 64$; $m_\pi = 300$ MeV; $m_\pi L = 4.08$ using a stout-smeared \cite{Morningstar:2003gk} Möbius \cite{Brower:2012vk} action for the charm, with Shamir \cite{Kaplan:1992bt, Shamir:1993zy, Furman:1995ky, Blum:1996jf, Blum:1997mz} action for the light and strange quarks. This is the first ensemble from what will be our production data set, and the bare charm quark mass is chosen to be $a m_h = 0.477$. We present results for 128 configurations, computing inversions on a single timeslice per configuration.


\subsection{Fitting the two-point correlation function}

\begin{figure}
\begin{center}
\includegraphics{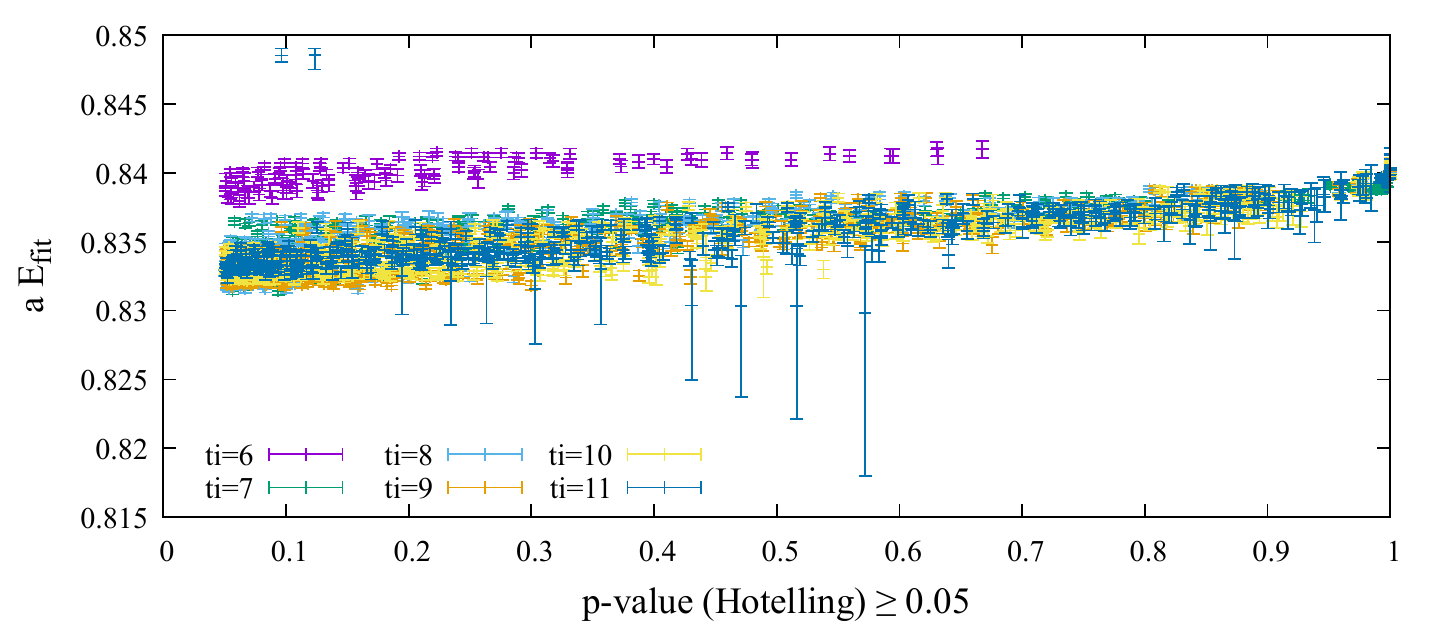}\\
\end{center}
\caption{Fit results for $a E_D \eval_{a m_h=0.477}$ extracted from simultaneous, two-state fits to $C^{\lr{2}}_{PP}$ and $C^{\lr{2}}_{WP}$ at $\abs{\vb{n}}^2 = 2$, scanning over independent fit ranges for each correlator. The horizontal axis is the $p$-value of the fit using the Hotelling distribution and colour is used to indicate the initial fit time on $C^{\lr{2}}_{PP}$ (all other labels suppressed).} \label{fig:fit_E0_D_s}
\end{figure}

In order to produce the mixed correlators defined in equations \eqref{eq:mixed_2pt} and \eqref{eq:mixed_3pt}, we extract the overlap coefficients per \eqref{eq:corr_2pt} from simultaneous, two-state fits to $C^{\lr{2}}_{PP}$ and $C^{\lr{2}}_{WP}$ (see fig~\ref{fig:fit_E0_D_s}).

Fit results are stable and consistent over the scanned fit ranges, except where fit ranges start very early or very late. Prior results \cite{Boyle:2018knm} (Table 6) bracket the values determined here:\\
$a m_D \eval_{a m_h=0.50} = 0.8489(11)$ and $a m_D \eval_{a m_h=0.41} = 0.74931(91)$.


\subsection{Extracting $Z_V$} \label{sec:Z_V}

\begin{figure}
\begin{center}
\includegraphics{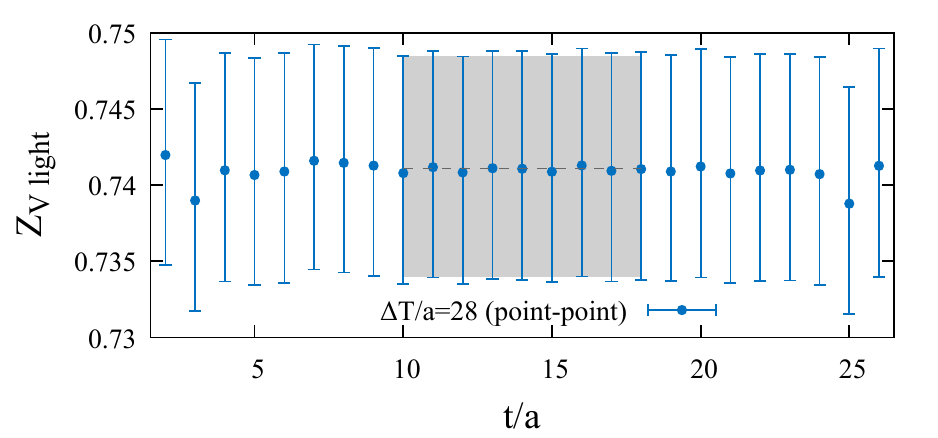}\\
\end{center}
\caption{$Z_{V \, \textrm{light}}^K$ extracted from Kaon 2- and 3-point functions with $\flatfrac{\Delta T}{a} = 28$. A correlated fit over $\flatfrac{t}{a} \in \lrb{10,18}$ (shaded region) yields $Z_V^K = 0.741(7)$, compatible with the published result $0.74563(13)$ \protect\cite{Boyle:2015hfa}.} \label{fig:fit_Z_V}
\end{figure}

Due to Lorentz invariance, the renormalised matrix element is parameterised in terms of the form factors $f_+$ and $f_-$ as
\begin{align}
Z_V \mel{P_f \lr{\vb{p}_f}}{V^\mu (\vb{q}^2)}{P_i \lr{\vb{p}_i}} &= f_+^{P_i P_f} (q^2) \lr{p_i + p_f}^\mu + f_-^{P_i P_f} (q^2) \lr{p_i - p_f}^\mu.
\end{align}
Due to charge conservation, the form factor $f_+$ at vanishing momentum transfer is unity, i.e. $f_+^{P_i P_i} = 1$. Considering the rest frame where $p_i = p_f \implies p_i + p_f = \lr{2 E_i, \vb{0}}$ and $p_i - p_f = 0$ we find (in the limit as $ \Delta T \gg t \gg 0$)
\begin{align}
\label{eq:ZV_renorm} Z_V &= \frac{2 E_i}{\mel{P_i \lr{\vb{0}}}{V_0 \lr{\vb{0}}}{P_i \lr{\vb{0}}}} \simeq \frac{C^{\lr{2}}_{ii} \lr{\Delta T, \vb{0}}}{C_{ii}^{4, \textrm{bare}} \lr{\Delta T, t, \vb{0}, \vb{0}}} .
\end{align}
We use this as a prescription to extract $Z_V$, emphasising that the denominator in \eqref{eq:ZV_renorm} is the bare (unrenormalised) correlator, following \cite{Boyle:2013}. $Z_V$ is extracted by forming the ratio defined on the right hand side of equation \eqref{eq:ZV_renorm} and fitting it to a constant in the region where excited state contributions are negligible (see fig~\ref{fig:fit_Z_V}).

We have a different action for the heavy vs the strange and light quark propagators and as a temporary measure we treat the mixed action bilinear using $Z_{V \, \textrm{mixed}} = \sqrt{Z_{V \, \textrm{heavy}} Z_{V \, \textrm{light/strange}}}$. However, in future we plan to use non-perturbative renormalisation \cite{Boyle:2017aa} on the mixed action bilinear.


\subsection{Three-point data}

Our aim is to map out the $q^2$ dependence of the form factors over the entire physical kinematic range. In our setup, the heavier meson is always kept at rest and the integer momentum $\vb{n} = \vb{p} \flatfrac{L}{\lr{2 \pi}}$ of the lighter meson labels the process. The momentum transfer to the lepton pair is $q = p_i - p_f$ (see fig~\ref{fig:continuum_decay}), for which the physical range is $m_\ell^2 \le q^2 \le q^2_{\textrm{max}} = (M_{P_f} - M_{P_i})^2$. For $\vb{n} = \vb{0}$, $q^2 = q^2_{\textrm{max}}$ and as $\abs{\vb{n}}^2$ increases we move down in $q^2$, reaching $q^2 \simeq 0$ for $\abs{\vb{n}}^2=4$ on the $M1$ ensemble for the mesons of interest.

In order to assess their statistical properties, we produced both ratios defined in equation \eqref{eq:R_alpha} for all three decays: $D_s \rightarrow K$, $D \rightarrow K$ and $D \rightarrow \pi$. The denominator for $R_2$ requires production of additional three-point correlators, making the $R_2$ ratio twice as expensive to calculate. We found that the results for $R_1$ and $R_2$ are consistent, with errors of the same magnitude. Due to the reduced cost of data production, we anticipate that we will only generate data for $R_1$ in the future. For this reason, only data for $R_1$ is shown in figures~\ref{fig:3pt_n_0}-\ref{fig:3pt_PW}.

These proceedings show our methodology and give a flavour of the quality of our data produced to date. As we are still in the process of refining the analysis strategy, the data shown are
preliminary.


\subsubsection{Temporal component of $R_1$ at $q^2_{\textrm{max}}$}

\begin{figure}
	\centering
	\begin{subfigure}[t]{0.47\textwidth}
		\centering
		\includegraphics{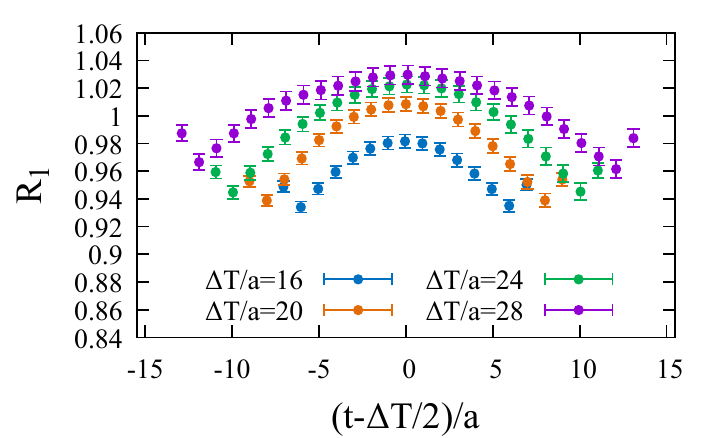}
		\caption{~$R_1^4 \lr{q^2_{\textrm{max}}}$, $D_s \rightarrow K$} \label{fig:R1_Ds_K}
	\end{subfigure}
	\begin{subfigure}[t]{0.47\textwidth}
		\centering
		\includegraphics{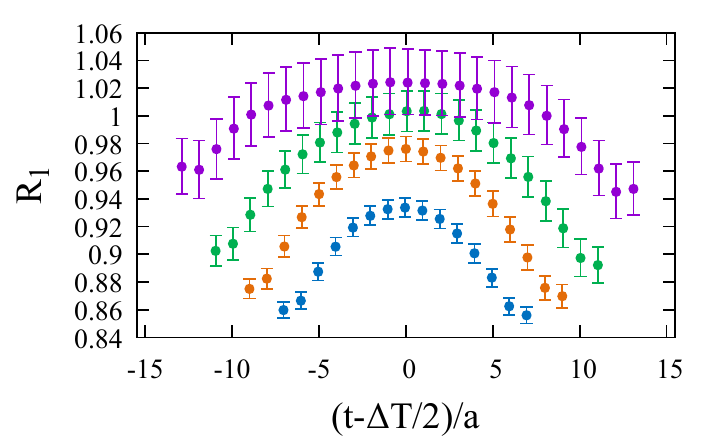}
		\caption{~$R_1^4 \lr{q^2_{\textrm{max}}}$, $D \rightarrow \pi$} \label{fig:R1_D_pi}
	\end{subfigure}
	\caption{Temporal component of $R_1$ for $D_s \rightarrow K$ and $D \rightarrow \pi$ decays for $\abs{\vb{n}}^2=0$, $V^0 \lr{q^2_{\textrm{max}}}$ (point source and point sink). Excited state contamination is clearly present for small $\flatfrac{\Delta T}{a}$ (as evidenced by the lack of plateau and failure to agree with other source-sink separations). At larger $\flatfrac{\Delta T}{a}$ the data overlap and we see a plateau, but it is not clear that excited states have been fully eliminated. A fit to multiple $\flatfrac{\Delta T}{a}$ will likely be required to reliably obtain the asymptotic plateau with small error.} \label{fig:3pt_n_0}
\end{figure}

The $D_s \rightarrow K$ decay is a charm to light transition with a strange spectator, whereas the $D \rightarrow \pi$ decay has a light spectator. The light propagator on the lattice is noisier than the strange, resulting in smaller relative uncertainties in fig~\ref{fig:R1_Ds_K} compared with fig~\ref{fig:R1_D_pi}.

There is clearly excited state contamination at small $\flatfrac{\Delta T}{a}$, which appears much reduced at larger $\flatfrac{\Delta T}{a}$ although not necessarily fully suppressed. As expected, errors grow quite distinctly with $\flatfrac{\Delta T}{a}$. We would ideally like to use the statistically cleaner data from smaller $\flatfrac{\Delta T}{a}$ in our analysis. This means we will need to model the excited state behaviour and include that in our fits for our full analysis -- the double-ratios alone will not be sufficient to fully control the contamination from the excited states.


\vspace{-0.05cm} 
\subsubsection{Spatial and temporal component of $R_1$ at non-zero momentum}

\begin{figure}
	\centering
	\begin{subfigure}[t]{0.47\textwidth}
		\centering
		\includegraphics{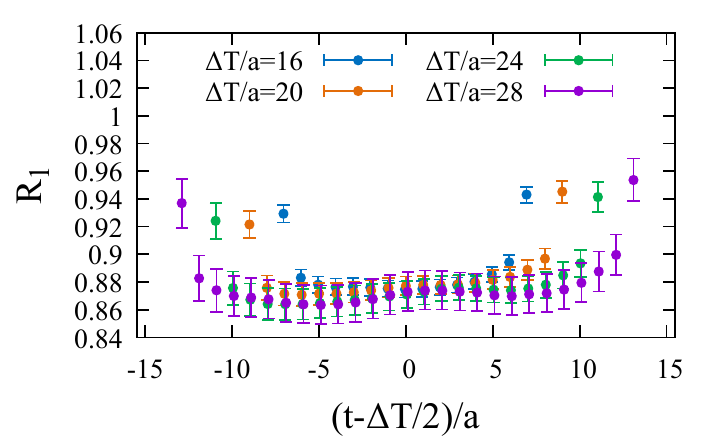}
		\caption{~$R_1^4 \lr{n^2 = 1}$} \label{fig:R1_Ds_K_p2_1_gT}
	\end{subfigure}
	\begin{subfigure}[t]{0.47\textwidth}
		\centering
		\includegraphics{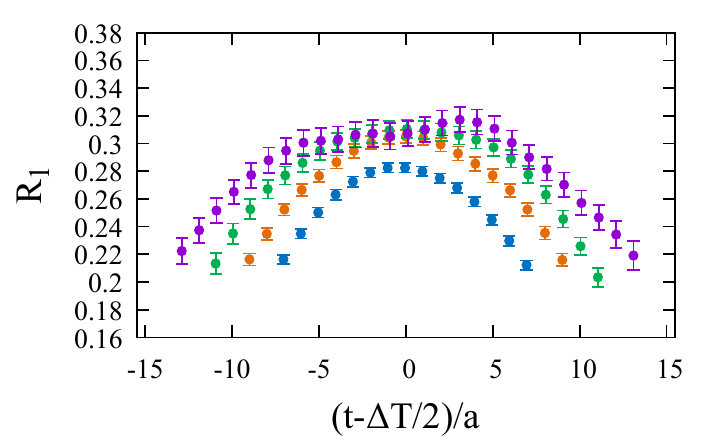}
		\caption{~$R_1^{i} \lr{n^2 = 1}$} \label{fig:R1_Ds_K_p2_1_gXYZ}
	\end{subfigure}
	\caption{Temporal (fig~\protect\ref{fig:R1_Ds_K_p2_1_gT}) and spatial (fig~\protect\ref{fig:R1_Ds_K_p2_1_gXYZ}) components of $R_1$ for $D_s \rightarrow K$ decays for $\abs{\vb{n}}^2=1$, (point-point).} \label{fig:3pt_n_1}
\end{figure}

Data have been produced across the physically allowable kinematic range. Results for spatial and temporal components of the vector current with one unit of momentum are shown in fig~\ref{fig:3pt_n_1}.

The errors are larger than for $q^2_{\textrm{max}}$, but still under control. Even though all but the smallest source-sink separations reach a consistent plateau value within statistical uncertainties, an analysis taking the excited states into account will allow us to quantify residual excited state contamination and thereby utilise the most precise data points whilst maintaining control over systematic uncertainties.

\subsubsection{Data for $R_1$ near $q^2 = 0$}

\begin{figure}
	\centering
	\begin{subfigure}[t]{0.47\textwidth}
		\centering
		\includegraphics{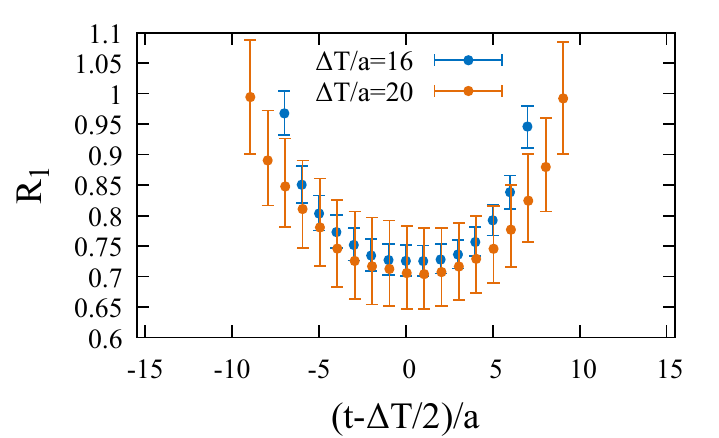}
		\caption{~$R_1^4 \lr{\sim q_0}$} \label{fig:R1_Ds_K_p2_4_gT}
	\end{subfigure}
	\begin{subfigure}[t]{0.47\textwidth}
		\centering
		\includegraphics{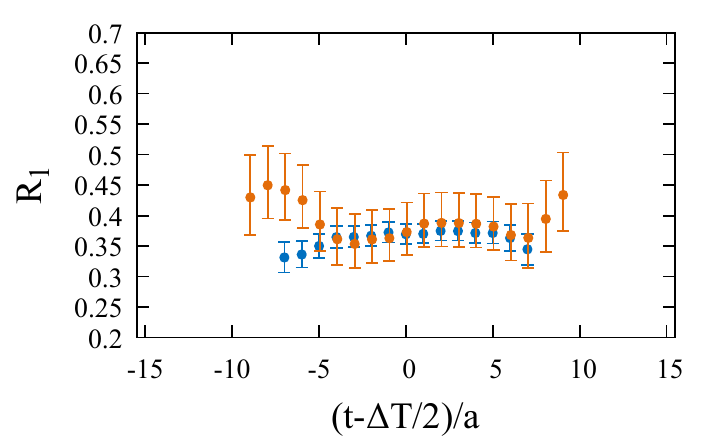}
		\caption{~$R_1^{i} \lr{\sim q_0}$} \label{fig:R1_Ds_K_p2_4_gXYZ}
	\end{subfigure}
	\caption{Temporal (fig~\protect\ref{fig:R1_Ds_K_p2_4_gT}) and spatial (fig~\protect\ref{fig:R1_Ds_K_p2_4_gXYZ}) components of $R_1$ for $D_s \rightarrow K$ decays near $q_0$ (i.e. $\abs{\vb{n}}^2=4$), (point-point).} \label{fig:3pt_n_4}
\end{figure}

Near the maximum recoil point $q^2 = 0$, the final state meson carries several units of momentum, resulting in larger statistical uncertainties.
For visibility, we have removed the very noisy data points corresponding to source-sink separations of $\flatfrac{\Delta T}{a} \ge 24$ from figure~\ref{fig:3pt_n_4}.

We are currently investigating whether data should be produced over a smaller range of $\flatfrac{\Delta T}{a}$ in finer increments as part of our fitting strategy evaluation. We might also increase statistics.


\subsubsection{Fitting strategy}

\begin{figure}
	\centering
	\begin{subfigure}[t]{0.47\textwidth}
		\centering
		\includegraphics{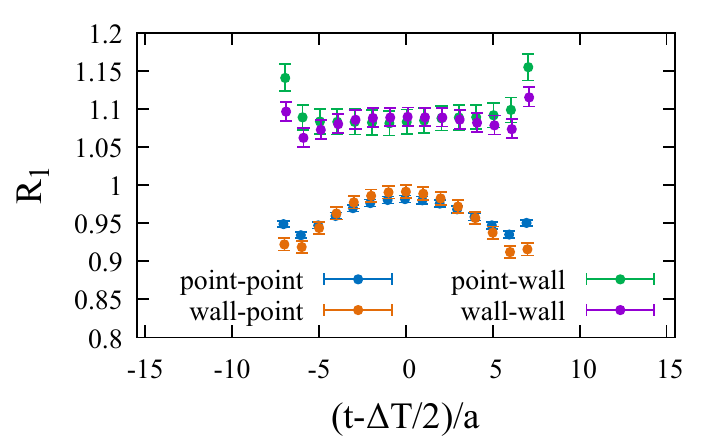}
		\caption{~$R_1^4 \lr{q^2_{\textrm{max}}}$ ($\flatfrac{\Delta T}{a} = 16$)} \label{fig:R1_Ds_K_p2_0_DT_16}
	\end{subfigure}
	\begin{subfigure}[t]{0.47\textwidth}
		\centering
		\includegraphics{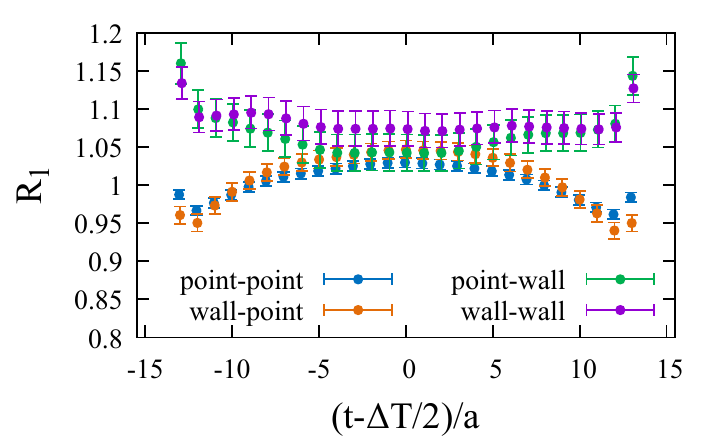}
		\caption{~$R_1^4 \lr{q^2_{\textrm{max}}}$ ($\flatfrac{\Delta T}{a} = 28$)} \label{fig:R1_Ds_K_p2_0_DT_28}
	\end{subfigure}
	\caption{Point(wall)-sink temporal components of $R_1$ for $D_s \rightarrow K$ approach plateau from below(above). This is consistent across source-sink separations,
	e.g.: $\flatfrac{\Delta T}{a} = 16$ (fig~\protect\ref{fig:R1_Ds_K_p2_0_DT_16}) and $\flatfrac{\Delta T}{a} = 28$ (fig~\protect\ref{fig:R1_Ds_K_p2_0_DT_28}).} \label{fig:3pt_PW}
\end{figure}

Figures~\ref{fig:3pt_n_0}-\ref{fig:3pt_n_4} present data for the ratio $R_1$ for point sources and sinks. In addition, we also produced data with wall sources and sinks as well as the mixed cases. Figure~\ref{fig:3pt_PW} shows this data at fixed source-sink separation for the temporal component of $R_1$ at zero recoil ($q^2_{\textrm{max}}$). We observe the qualitatively different approach to the plateau of the wall-sink data, which approaches the plateau from above, while the point data approaches the plateau from below. We intend to utilise the various features of our data set by simultaneously fitting the different operator choices and multiple source-sink separations.

Based on the data we have, we expect to be able to extract the form factors over the entire physical kinematic range.


\section{Summary and Outlook}

We performed a study on point-wall diagonalisation as a method for reducing excited state contamination. The method works very well for two-point functions, however, extending this to the three-point functions is still work in progress.

We have produced two- and three-point correlation functions for $D_{\lr{s}} \rightarrow K / \pi$ semileptonic decays on our first ensemble, which we present here. After examining the three-point correlation function data on this first ensemble, we conclude that we will need to use wall separations which do not fully eliminate excited state contamination (even when analysed using double-ratio methods). We will, therefore, simultaneously fit correlator data from multiple wall separations, giving greater control over excited state contamination.

Having both point and wall correlation function data, with differing approaches to plateau, gives us extra control over excited states. We can include the point and wall correlation functions separately in our simultaneous fits and/or use a rotated basis for our fits. Due to noise issues with the wall sink data, we may use a one-sided diagonalisation at the source only for the three-point correlation function -- but this is to be determined and work is ongoing to finalise our analysis.

Since the lattice conference, we have optimised our performance on DiRAC's new Tursa NVidia A100 GPU-based supercomputer and we have produced data for a second ensemble.

The target result is the $q^2$-dependence of the $D_{\lr{s}} \rightarrow K / \pi$ form factors over the entire physical $q^2$ range, and results from our first ensemble indicate that percent-scale errors are achievable.


\acknowledgments

{\small{Kind thanks to the RBC/UKQCD collaboration for many invaluable discussions and helpful suggestions. Results presented here were produced using Grid \cite{Grid:2016} and Hadrons \cite{Hadrons:2020}.

This work used the DiRAC Extreme Scaling service at the University of Edinburgh, operated by the Edinburgh Parallel Computing Centre on behalf of the STFC DiRAC HPC Facility (\url{https://www.dirac.ac.uk}). This equipment was funded by BEIS capital funding via STFC capital grant ST/R00238X/1 and STFC DiRAC Operations grant ST/R001006/1. DiRAC is part of the National e-Infrastructure.}}

P.B. has been supported in part by the U.S. Department of Energy, Office of Science, Office of Nuclear Physics under the Contract No. DE-SC-0012704 (BNL). P.B. has also received support from the Royal Society Wolfson Research Merit award WM/60035.
L.D.D. is supported by the U.K. Science and Technology Facility Council (STFC) grant ST/P000630/1.
F.E. and A.P. are supported in part by UK STFC grant ST/P000630/1. F.E. and A.P. also received funding from the European Research Council (ERC) under the European Union’s Horizon 2020 research and innovation programme under grant agreement No 757646 and A.P. additionally under grant agreement No 813942.
A.J. and J.F. acknowledge funding from STFC consolidated grant ST/P000711/1 and A.J. from ST/T000775/1.
M.M. gratefully acknowledges support from the STFC in the form of a fully funded PhD studentship.
J.T.T.: the project leading to this application has received funding from the European Union's Horizon 2020 research and innovation programme under the Marie Sk{\l}odowska-Curie grant agreement No 894103.

\bibliographystyle{JHEP}
\bibliography{Lat21Refs.bib}

\end{document}